\documentclass{aa}

\usepackage{graphicx}
\usepackage{subcaption}
\usepackage{txfonts}
\usepackage{natbib}
\usepackage{amsmath}
\usepackage{multicol}
\usepackage{stfloats}
\usepackage{color}
\usepackage{yfonts}
\usepackage{xcolor}
\usepackage{ulem}

\usepackage[colorlinks=true, linkcolor=blue, citecolor=blue, filecolor=blue, urlcolor=blue]{hyperref}
\begin{document} 
   \title{Sub-second cadence structure of optical flares on AD\,Leo}

  \author{B.~Schmercz 
          \inst{1,2,3}
        \and
          B.~Seli
          \inst{1,2,3}
        \and
          K.~Vida
          \inst{1,2}
        \and
          L.~Kriskovics
          \inst{1,2}
        \and
          A.~G\"orgei
          \inst{1,2,3}
        \and
          K.~Ol\'ah
          \inst{1,2}
        \and
          Zs.~Reg\'aly
          \inst{1,2}
}
   \institute{Konkoly Observatory, HUN-REN Research Centre for Astronomy and Earth Sciences, Konkoly Thege \'ut 15-17., H-1121 Budapest, Hungary\\
   	\email{schmercz.blanka@csfk.org}
        \and
            HUN-REN RCAES, MTA Centre of Excellence, Budapest, Konkoly Thege út 15-17., H-1121 Budapest, Hungary
        \and
            E\"otv\"os University, Department of Astronomy, Pf. 32, 1518 Budapest, Hungary 
}             
   \date{Received ...; accepted ...}

 \abstract
{Stellar flares are sudden brightenings caused by magnetic reconnection and are frequently observed on late-type stars. High-cadence photometry of flares provides valuable insights into the mechanisms of these events, yet such observations remain scarce.} 
  %leave it empty if necessary  
{We seek to explore the sub-second fine structure of stellar flares and assess the information content in high-speed photometry.}
{New 0.3\,s-cadence photometry from a six-year-long observing campaign of the active M-dwarf AD\,Leo is presented. We use time--frequency analysis to detect quasi-periodic pulsations in the decay phase of flares. We explore statistical measures of time series complexity of the detected flares to quantify the information gain achievable with high-cadence photometry. }
{We detect 42 flares in 211\,hours of observations. The flare frequency distribution is consistent with the previous literature. We find no quasi-periodic pulsations with periods below a few seconds, and identify two candidate signals with periods around 1 and 3\,min. Using different measures of complexity on the binned flare light curves we confirm the advantages of high observing cadence. However, we also find a plateau up to a binning of $\approx$4--5\,s for a few complex flares, suggesting that an exposure time of a few seconds is usually enough to retain most of the information carried by a single-filter observation.}
{New photometric observations of AD\,Leo revealed sub-structures of flare light curves on the timescale of a few seconds, but we found no features on timescales below that.}

   \maketitle

\section{Introduction}

Stellar flares are caused by the sudden release of magnetic energy in stellar atmospheres, observable as an increase of stellar luminosity for minutes to hours \citep{2024LRSP...21....1K}. With the advent of space-borne photometry, hundreds of thousands of flares have been observed with a few minute cadence \citep[e.g.,][]{2019ApJS..241...29Y, 2022ApJ...925L...9F, 2024AJ....168..234L, 2025A&A...694A.161S}. However, flare observations with sub-second time resolution are relatively rare.
One of the early experiments was the work of \cite{1988SvAL...14...65B} who used a 6-m telescope equipped with the \textit{Multichannel Analysis of Nanosecond Intensity Alterations} (MANIA) instrument, capable of detecting individual photon arrival times with an accuracy of 0.3\,$\mu$s. They detected 118 flares on 8 stars in 35\,hours of observations. They found no fine structure on timescales shorter than 0.1\,s, and only detected statistically significant variations on timescales longer than 0.5\,s. More details on these observations were published by \cite{1988BCrAO..79...67B}, including light curves showing flares with different morphologies, such as double peaks and possible oscillations. Most of these events are shorter than 1\,min. High-cadence observations also enable the detection of quasi-periodic pulsations \citep[QPPs,][]{2021SSRv..217...66Z}. \cite{2024arXiv241207580P} used the MANIA instrument to detect 157 flares on M dwarfs in 70\,hours of observations. The 1\,s time resolution allowed them to find 13 QPPs, with periods ranging from 6 to 107\,s. High-cadence observations have also been carried out by \cite{2022PASJ...74.1069A}, who observed 22 flares on M dwarfs with 1\,s cadence. They note the presence of flat peaks, which are rarely resolved with longer cadence, and provide a new analytical flare template. 

A few studies focused on detecting flares with the shortest possible duration, analogous to solar micro- and nanoflares. \cite{1995ApJ...451..795R} used the Hubble Space Telescope to observe CN\,Leo in the ultraviolet, finding 32 flare-like events, most of which lasted for less than 10\,s. \cite{1997RMxAA..33..107T} reported impulsive events with a duration of less than 1\,s, mostly following longer flares. \cite{2016A&A...589A..48S} observed UV\,Cet with 4\,$\mu$s time resolution and found two flares with rise and decay times of 2\,s.

AD\,Leo (GJ\,388) is one of the brightest and most active flare stars, making it an ideal target in a hunt for stellar flares. It has a spectral type of M3V, a rotational period of $2.227$\,days, and it is probably viewed pole-on \citep{2022A&A...666A.143K}. AD\,Leo is a prime target to search for spectral signatures of flares \citep{1991ApJ...378..725H, 2020PASJ...72...68N, 2020A&A...637A..13M} and radio bursts \citep{2006ApJ...637.1016O, 2023ApJ...953...65Z, 2024A&A...686A..51M}. The evolution of its magnetic field has been followed with spectropolarimetric observations \citep{2008MNRAS.390..567M, 2018MNRAS.479.4836L, 2023A&A...676A..56B}. AD\,Leo has a long history of photometric observations of its flare activity \citep[e.g.,][among many others]{1959PASP...71..517A, 1969IBVS..345....2S, 1984ApJS...54..375P, 2003ApJ...597..535H, 2010AJ....140..483D}, including multicolor fast photometric observations \citep{2013ARep...57..603L} and more recent space photometric studies with MOST \citep{2012PASP..124..545H}, TESS \citep{2023PASP..135f4201B, 2025ApJ...980..196R}, and CHEOPS \citep{2024A&A...686A.239B}.

In this paper, we present new fast photometric observations of AD\,Leo and aim to identify fine structure in stellar flare light curves, in order to find the shortest timescale where any kind of variation can be expected.

\section{Data}

Observations were carried out at Piszkéstető Mountain Station, Konkoly Observatory, using the 1-meter Ritchey--Chrétien--Coudé telescope ($D = 1.02$\,m, $f = 13.8$\,m), which provides a field of view of approximately 3.5\,arcminutes. 

Observations were performed with OCELOT (Optical Camera with ELectronmultiplying OutpuT), an Andor iXon+888 back-illuminated \mbox{EMCCD} (Electron Multiplying CCD) camera in a custom casing for temperature control, using Johnson $B$ filter, with a cadence of 0.286\,s, operating at $-80^\circ$C. An \mbox{EMCCD} camera is capable of achieving fast sampling with negligible readout time in frame transfer mode, and it can reliably detect photons even in low light conditions \citep{2004SPIE.5499..162S}. We selected the shortest available exposure time for the 16-bit mode with $4 \times 4$ binning that still allowed the field of view needed to include a comparison star. Measurements were carried out on 49 nights between 2019 February 23, and 2025 April 4 (see the full observing log in Table~\ref{table:obs_log}). In total, approximately 211 hours of observational data were collected.

Data reduction was performed using differential aperture photometry routines implemented with \texttt{photutils} \citep{larry_bradley_2025_14889440} and \texttt{astropy} \citep{2022ApJ...935..167A}, following dark subtraction and flat-field correction. BD+20\,2464 was used as the comparison star, which has a magnitude of 11.67 in the $B$ filter \citep{2000A&A...355L..27H}.

The noise level of the light curve is generally high (0.05\,mag on average, see Table~\ref{table:obs_log}), and it also depends on the sky condition. This is partly a consequence of the limited signal transmitted through the Johnson $B$ filter, which we used to maximize the flare signal \citep[see e.g.,][]{2025A&A...699A..90B}. For visual aid in the figures, we enhance the details with smoothing.

\section{Methods}

\subsection{Flare energies}

Flare energies were estimated using the method described in \citet{2022A&A...668A.101O}. This approach is based on the concept of equivalent duration \citep[ED,][]{1972Ap&SS..19...75G}, which represents the time that a star in its quiescent state would require to emit the same amount of energy that is released during a flare. The ED is calculated as the integral of the relative flare intensity over its duration:
\begin{equation}
    \mathrm{ED} = \int_{t_1}^{t_2} \left( \frac{I(t)}{I_0} - 1 \right) dt,
\end{equation}
where $I(t)$ is the observed stellar intensity as a function of time, and $I_0$ denotes the quiescent (non-flaring) intensity level. To estimate the uncertainties of the EDs, we slightly varied the integration limits and examined the resulting changes.

Flare energies were then calculated by multiplying the EDs by the quiescent $B$-band luminosity ($L_B$). To estimate $L_B$ we used a BT-NextGen model spectrum \citep{1999ApJ...512..377H} for AD\,Leo with $T_\mathrm{eff} = 3400$\,K, $\log{g}=4.5$ and [Fe/H]=0, and the $B$ filter transmission curve. We determined the fraction of the model spectrum transmitted through the filter and multiplied it by the bolometric luminosity of the star from \cite{2000A&A...354.1021F} to derive the quiescent $B$-band luminosity of $L_B = 1.23 \times 10^{30}$\,erg\,s$^{-1}$.

To make the flare energies easier to compare with previous studies, the $B$-band energies were converted to bolometric values by assuming a blackbody flare temperature of 12~000\,K \citep{2023PASP..135f4201B}. As there is a large scatter in measured flare temperatures \citep[see, e.g.,][]{2022AJ....164..223R}, we estimate the uncertainty of the conversion by assuming a standard deviation of 3000\,K. Integrating a black body flare spectrum with the $B$ filter transmission curve results in a ratio of the bolometric to $B$-band energies of $7.69 \pm 1.78$, thus giving:
\begin{equation}\label{eq:energy_conversion}
    E_\mathrm{bol} = (7.69 \pm 1.78) \times L_B \times \mathrm{ED}
\end{equation}
If $dN$ represents the number of flares occurring within the energy interval $E + dE$, the flare energy distribution follows a power-law relation \citep[see e.g.,][]{2024LRSP...21....1K}:

\begin{equation}
    \frac{dN}{dE} \propto E^{-\alpha}.
\end{equation}
Integrating the relation gives the cumulative flare frequency distribution in logarithmic form:

\begin{equation}
    \log \nu = \beta \log E + a,
\end{equation}
where $\nu$ is the cumulative frequency of flares with energy above $E$, and $\beta = 1 - \alpha$. The index $\alpha$ characterizes the energy distribution: for $\alpha > 2$, low-energy flares dominate the total output; for $\alpha < 2$, high-energy flares contribute more \citep{1991SoPh..133..357H}.

\subsection{Identifying quasi-periodic pulsations}

To search for QPPs in complex flare light curves, different methods can be used in both the time and the frequency domain \citep{2019ApJS..244...44B}. Here we use a modified version of the Short-Term Fourier Transform (STFT).

In the STFT approach, the signal is multiplied by a Gaussian window centered at a given time, and the resulting segment is Fourier-transformed to extract the local frequency content. By sliding the window along the time axis, a time--frequency representation is produced, which can be visualized as a two-dimensional map when the window positions are sampled densely \citep{2009A&A...501..695K}.

In the modified version used here from the \texttt{seismolab} package \citep{2024JOSS....9.7118B}, the local frequency analysis is performed using the Lomb--Scargle periodogram instead of the standard Fourier transform. This approach is known as the Windowed Lomb--Scargle Transform. As these methods work best with stationary baselines, we first subtract the flare profile using a Savitzky--Golay filter \citep{1964AnaCh..36.1627S}. This smoothing algorithm fits low-order polynomials to successive local data segments, effectively reducing noise while preserving sharp features and peak structures. For the smoothing, we adopted a kernel size of 3\,min. Subsequently, for the windowed Lomb–Scargle transform of each flare, we applied a Gaussian window with a width of 1.2\,min.

The significance of potential signals was evaluated using a bootstrap procedure, similar to \cite{2026MNRAS.546S2256L}. We generated 10~000 surrogate light curves by randomly permuting the normalized residuals obtained after subtracting a smoothed version of the original light curve. The signal significance was then estimated from the distribution of the resulting maximum peak amplitudes. This method assumes white noise as the null hypothesis, so the significance we get can be overly optimistic if correlated noise is present.

\subsection{Measure of complexity}
\label{sect:complexity_methods}

High-cadence photometry can reveal details of flares that cannot be seen with longer integration times \citep[see e.g.,][]{2018ApJ...859...87Y, 2022ApJ...926..204H}. In this section we try to quantify this information gain by using different methods to measure the "complexity" of light curve segments containing flares. We aim to find a metric that is non-parametric (i.e., it does not assume a functional form for components), robust to random noise, and does not depend only on the number of points in the light curve segment, making the metric comparable across different flares. For example, we expect a single-peaked flare to have a low complexity score, while a flare with multiple peaks or QPPs should have a higher score, regardless of its amplitude and duration. We will use these techniques to rank the flares in the sample from least to most complex, and to see how the complexity of the flares change with binning, simulating different integration times. This way we can find an optimal integration time that maximizes the signal-to-noise ratio while still preserving the information content of the stellar flare light curve -- for the first time with real data.

\begin{figure}[tb]
    \centering
    \includegraphics[width=\columnwidth]{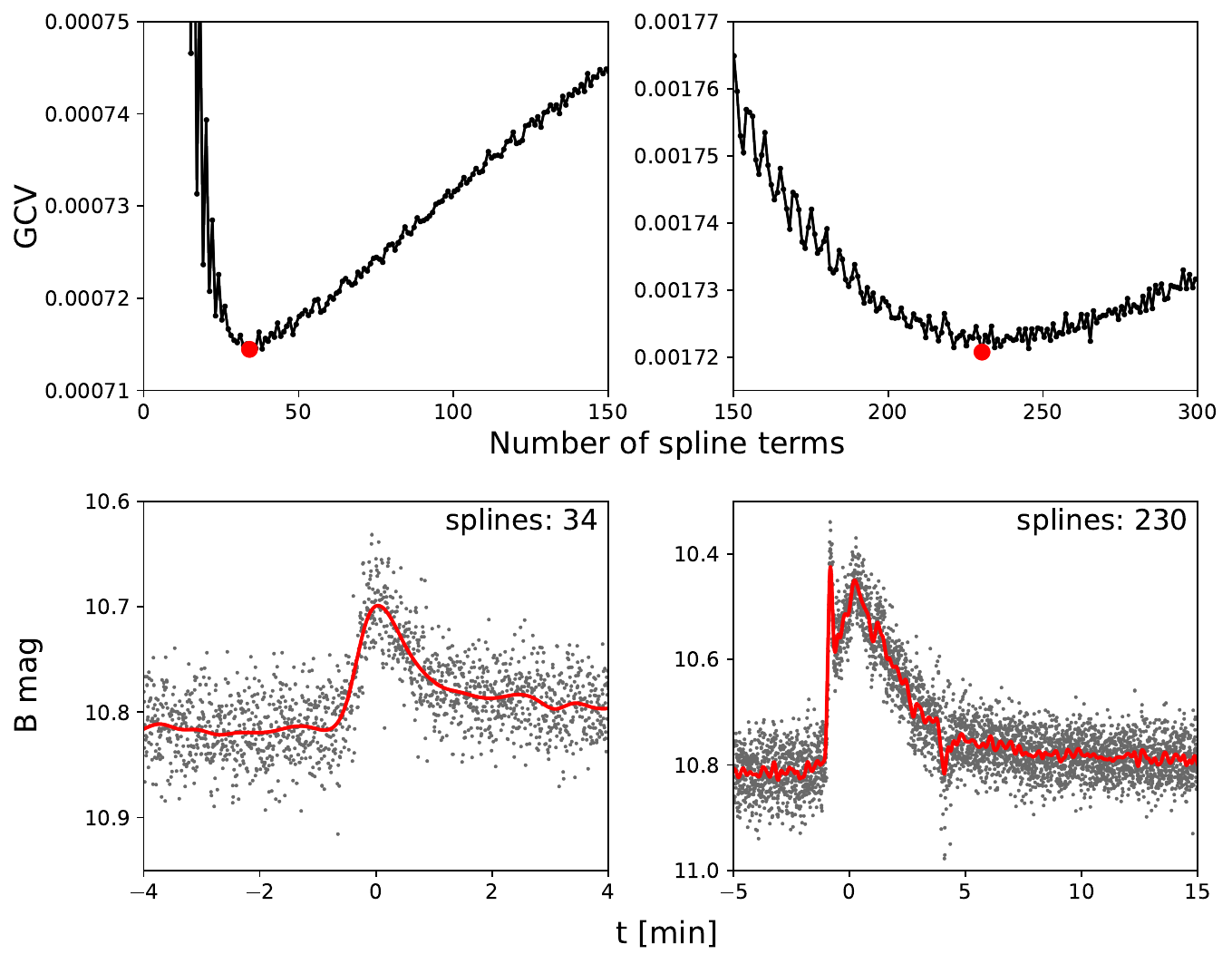}
    \caption{Complexity assessment with generalized additive models (GAMs) for two flares. Top: generalized cross-validation score as a function of the number of spline terms, with the red point marking the minimum. Bottom: flares with the best-fitting curve corresponding to this minimum; associated spline values are indicated in the panels.}
    \label{fig:gam_nsplines}
\end{figure}

We first assessed flare complexity  using generalized additive models
\citep[GAMs,][]{hastie1986} via \texttt{pyGAM} \citep{2018zndo...1208724S}, which extend linear models with smooth functions using penalized B-splines. The flexibility of these functions is controlled by the number of spline terms through the number and placement of knots, determining how finely non-linear patterns are captured. For each flare, the segment of the time series to be fitted was predefined, spanning from a few minutes before the flare onset to a few minutes after its end. Each flare was then fitted using a number of spline terms ranging from 4 to 400, while keeping all other model parameters at their default values. The resulting fits were evaluated using the generalized cross-validation (GCV) score, which estimates the model’s prediction error while balancing fit quality and model complexity. This helps to avoid overfitting by penalizing overly complex models. Since lower GCV values indicate better fits, the value corresponding to the minimum GCV was adopted as the measure of model complexity, as illustrated in Fig.~\ref{fig:gam_nsplines}. This method is easy to interpret and works well with noisy data, as GAMs are often applied for smoothing and interpolating noisy datasets \citep[see e.g.,][]{2023A&A...678A..44C, 2024SpWea..2203680B}.

The second method involves the effective degrees of freedom (EDF) from a GAM fit, computed with a fixed number of spline terms. The EDF depends on the number of GAM parameters and also on the smoothing penalty, so its value is in general not an integer. Using a fixed number of spline terms is necessary to compare the EDF across different events, but it risks underfitting for larger flares or overfitting for smaller ones. We used 100 spline terms, which gives a reasonable fit for all flares.

Increment entropy provides an alternative way to quantify dataset complexity \citep{2016Entrp..18...22L}. It measures the complexity of time series data by calculating the Shannon entropy of patterns derived from the increments (differences) between consecutive data points. The metric was computed using the \texttt{EntropyHub} toolkit \citep{2021PLoSO..1659448F}, which analyzes the distribution of increments between consecutive data points encoded as symbolic patterns. Higher values reflect greater variability and less predictable changes.

\section{Results and discussion}

\subsection{Flare detection}

A total of 42 flare events were identified through visual inspection of the light curve, as shown in Fig.~\ref{fig:all_flares} in the Appendix. For each flare, the light curve was fitted with the flare profile described in \cite{2014ApJ...797..122D}. The resulting parameters (peak time, amplitude, and $t_{1/2}$ full width at half maximum) are listed in Table~\ref{table:flares}, along with the three complexity metrics. The table also gives the corresponding $B$-band and bolometric flare energies, along with their uncertainties obtained by propagating the errors in the flare temperature and the EDs.

\subsection{Flare frequency distribution}

\begin{figure}[htb]
    \centering
    \includegraphics[width=\columnwidth]{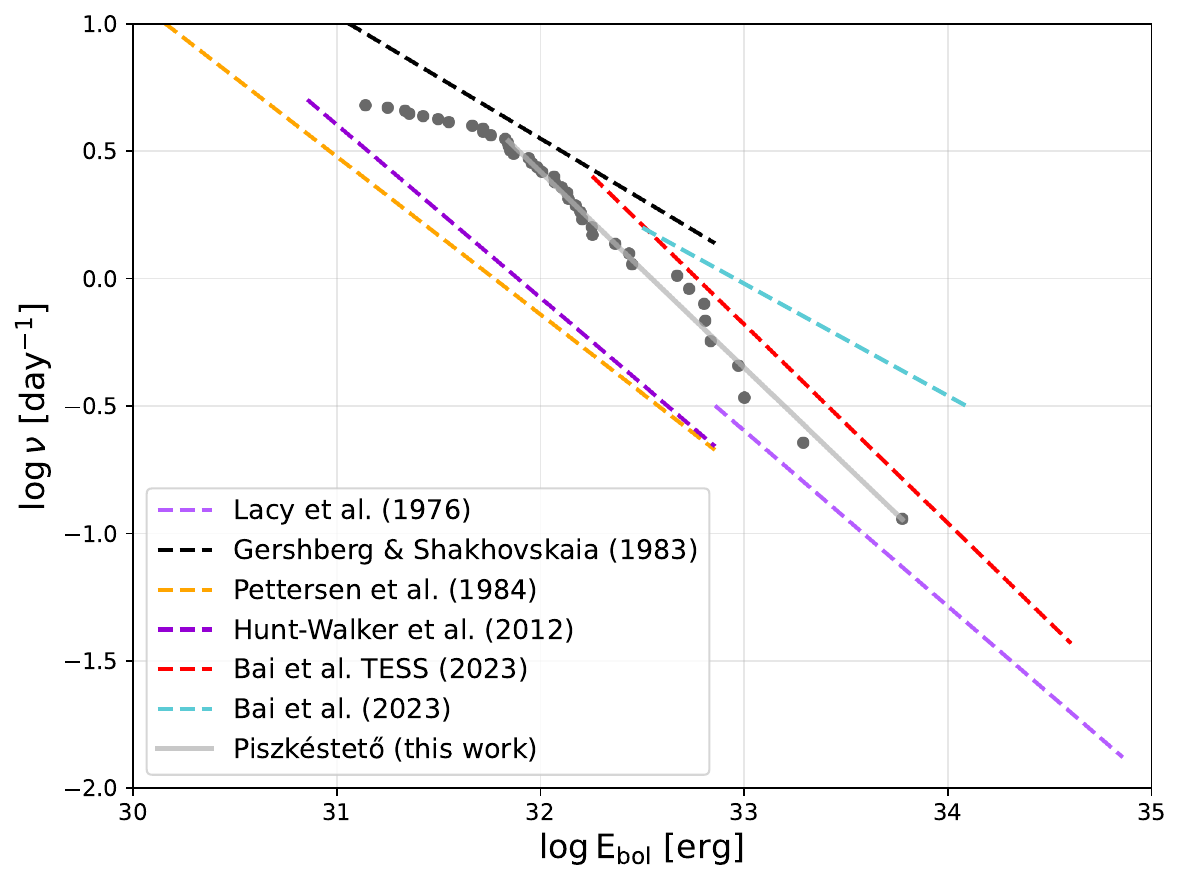}
    \caption{The flare frequency distribution derived in this work (grey data points and corresponding best-fit line), compared to fitted trends reported in the literature.}
    \label{fig:ffd}
\end{figure}

The flare frequency distribution (FFD) of AD\,Leo was determined, and the following power law was fitted for energies above $E_\mathrm{bol} = 6.8 \times 10^{31}$\,erg (corresponding to the minimum energy at which the $E_\mathrm{min}$–slope function flattens, similar to the method of \citealt{2024MNRAS.527.8290P}):

\begin{equation}
\log \nu = (-0.70 \pm 0.02) \log E_{\mathrm{bol}} + (22.5 \pm 0.5),
\end{equation}
resulting in $\alpha = 1.70 \pm 0.02$. This FFD, together with the fitted line is shown in Fig.~\ref{fig:ffd}. For context, the figure also includes previously published FFDs from the literature, with all energies converted to bolometric values for direct comparison. Potential temporal variations in the FFD are not addressed in this study due to the lack of sufficiently extensive data.

Earlier FFDs of AD Leo show a range of slopes, generally consistent with our result.
\citet{1976ApJS...30...85L} reported $\alpha = 1.82 \pm 0.27$, although it was based on just 9 flares observed over 21.5\,hours.
\citet{1983Ap&SS..95..235G}, with over 1000\,hours of monitoring and 54 flares, found a flatter slope of $\alpha = 1.48 \pm 0.06$.
\citet{1984ApJS...54..375P} observed the star over several years and reported $\alpha = 1.62 \pm 0.09$, finding no significant year-to-year variations.
\citet{2012PASP..124..545H} identified 19 flares in about 140\,hours of continuous photometry, yielding $\alpha = 1.68 \pm 0.16$.
More recently, \citet{2023PASP..135f4201B} found $\alpha = 1.78 \pm 0.09$ from TESS and $\alpha = 1.44 \pm 0.15$ from \mbox{GWAC-F30}, although the latter was based on only 9 events. In addition, \citet{2025ApJ...980..196R} compared nine different $\alpha$ values obtained from both photometric and spectroscopic observations. Here we limit our comparison to photometric results.

The efficiency of flare detection and the accuracy of flare energy estimation are both affected by the exposure time \citep{2018ApJ...859...87Y}. To investigate how information derived from flare observations, e.g., FFD, depends on exposure time, the dataset was rebinned using bin sizes ranging from 1.2\,s to 5\,min. A flare was considered detected if it contained at least five data points and its peak exceeded the quiescent level by more than three standard deviations. The quiescent level and its standard deviation for each bin size were estimated by repeatedly rebinning the data with shifted bin starting points. The EDs and the FFD were then computed for each bin size. The ED of individual flares, as well as the location of break point due to low completeness and the overall shape of the FFD, remained nearly constant across all bin sizes. However, with larger bins, fewer flares were detected, and around 2.5\,min the break point itself disappeared, leaving only the high-energy region. We note that this analysis assumed zero readout time. A non-negligible readout time could bias the ED of complex flares.

\subsection{Multi-peaked flares}

Complex flares exhibiting multiple peaks are often interpreted as sequences of rapidly successive simple flare events \citep[e.g.,][]{2014ApJ...797..122D, vidaInvestigatingMagneticActivity2016}, a phenomenon that is also well documented on the Sun. Possible explanations include homologous flares, in which successive eruptions originate within the same active region and involve similar magnetic structures \citep{1963QJRAS...4...62E}, and sympathetic flares, where activity in one region can trigger flares in another through large-scale magnetic connections  \citep{torokModelMagneticallyCoupled2011}. 
Both processes have been observed on the Sun \citep[e.g.,][]{2017A&A...601A..39P, 2025A&A...698A.185Y,  2025A&A...694A..74G}, but the limited spatial resolution of stellar observations prevents us from distinguishing between them. Several mechanisms have been proposed to account for this behavior \citep[e.g.,][]{1976ApJ...204..530M, 2001ApJ...559.1171W}, although the precise cause remains a subject of ongoing investigation.

\begin{figure}[tb!]
    \centering
    \includegraphics[width=\columnwidth]{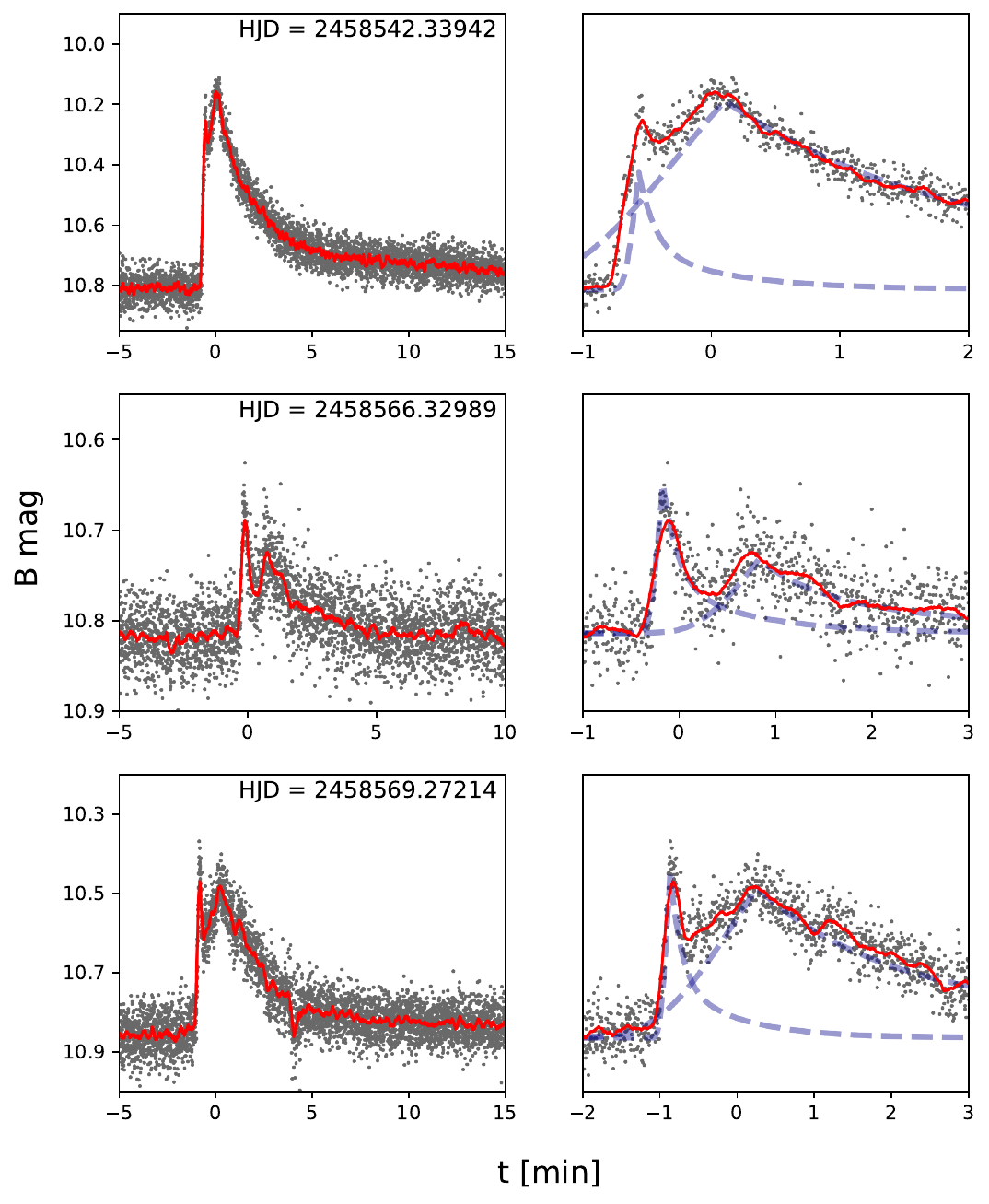}
    \caption{Light curves of flares exhibiting complex temporal structures are shown as grey points, with the corresponding smoothed curves indicated by red lines. The first column shows the full flare profiles, while the second column provides magnified views of the peaks, illustrating the two fitted flare components with dashed lines. The HJD indicated in each panel of the first column corresponds to t = 0 on the time axis.}
    \label{fig:multi_peaked_flares}
\end{figure}

In our sample, three flares displayed this complex behavior. The light curves of these complex flares are presented in Fig.~\ref{fig:multi_peaked_flares}, with the time difference between successive peaks being less than two minutes in each case. Considering that 42 flares were detected over 211 hours of observation, the probability that, in these three cases, the two peaks that represent separate flares occurred independently despite being so close together is very low. When each complex flare was decomposed into two events using the flare template of \cite{2014ApJ...797..122D}, the first event was generally shorter than the second, as reflected in its lower $t_{1/2}$ value, and for the flares at HJD = 2458566.32989 and HJD = 2458569.27214, the first event also reached a higher amplitude than the second. We also note that in these cases, the peak of the first component lies close to where the rise phase of the second component begins (see the right-hand panels of Fig.~\ref{fig:multi_peaked_flares}). Flares with similar morphologies often appear among complex flares, see e.g. Fig.~8 in \cite{2022PASJ...74.1069A}, Fig.~10 in \cite{2022ApJ...926..204H},
Fig.~5 in \cite{2024SoPh..299....8V},
or Fig.~1 in \cite{2025ApJ...984..186Y}.

\subsection{Quasi-periodic pulsations}

\begin{figure}[t!]
    \centering
    \includegraphics[width=\columnwidth]{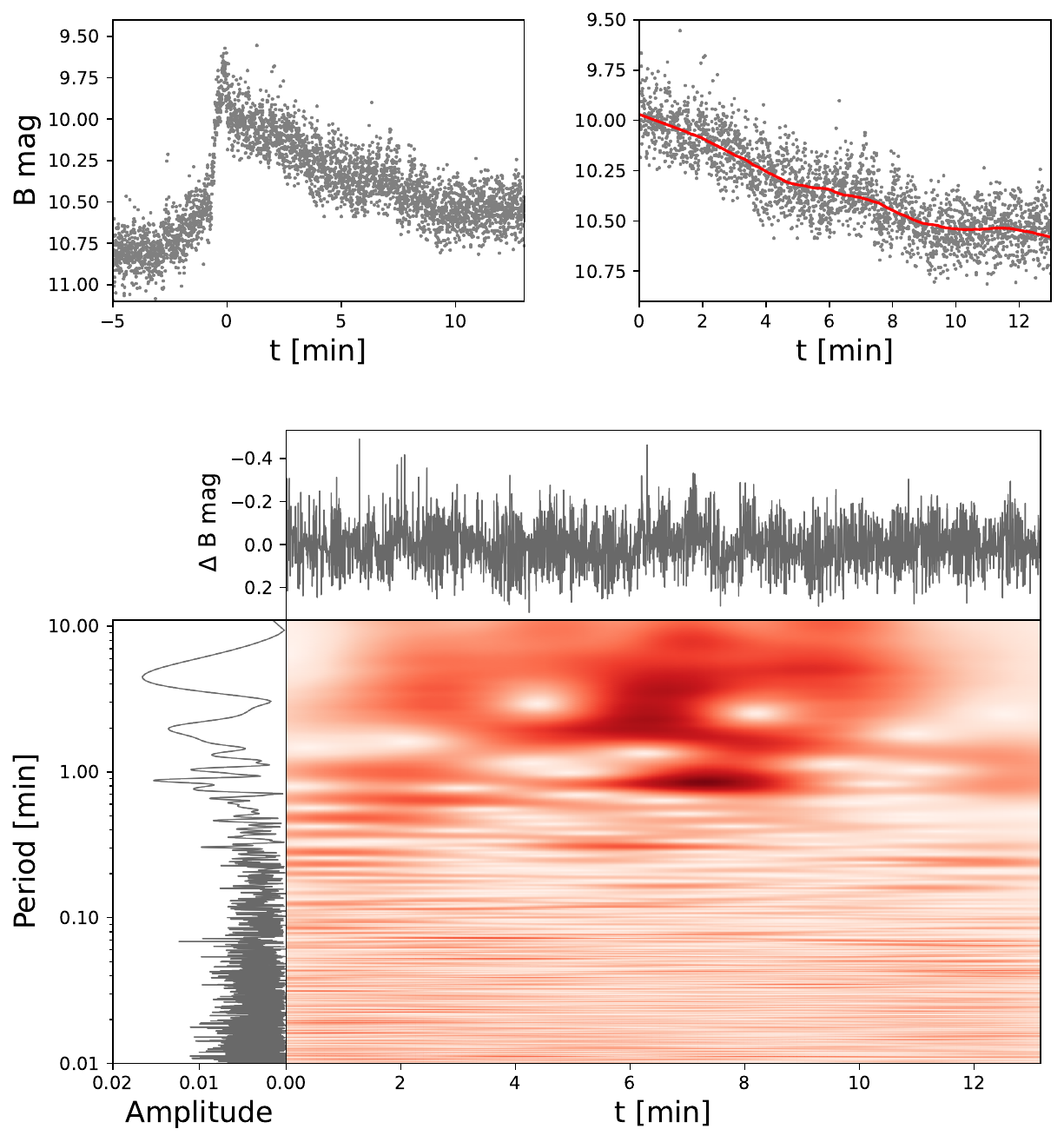}
    \caption{Time--frequency analysis of the flare at $\mathrm{HJD}=2458564.51777$. Red line in the upper right panel shows the smoothed light curve that was subtracted before the application of the windowed Lomb--Scargle periodogram. Possible quasi-periodic pulsation is present with a period around 1\,min (around $t=6-8$\,min).}
    \label{fig:stft_1}
\end{figure}

\begin{figure}[t!]
    \centering
    \includegraphics[width=\columnwidth]{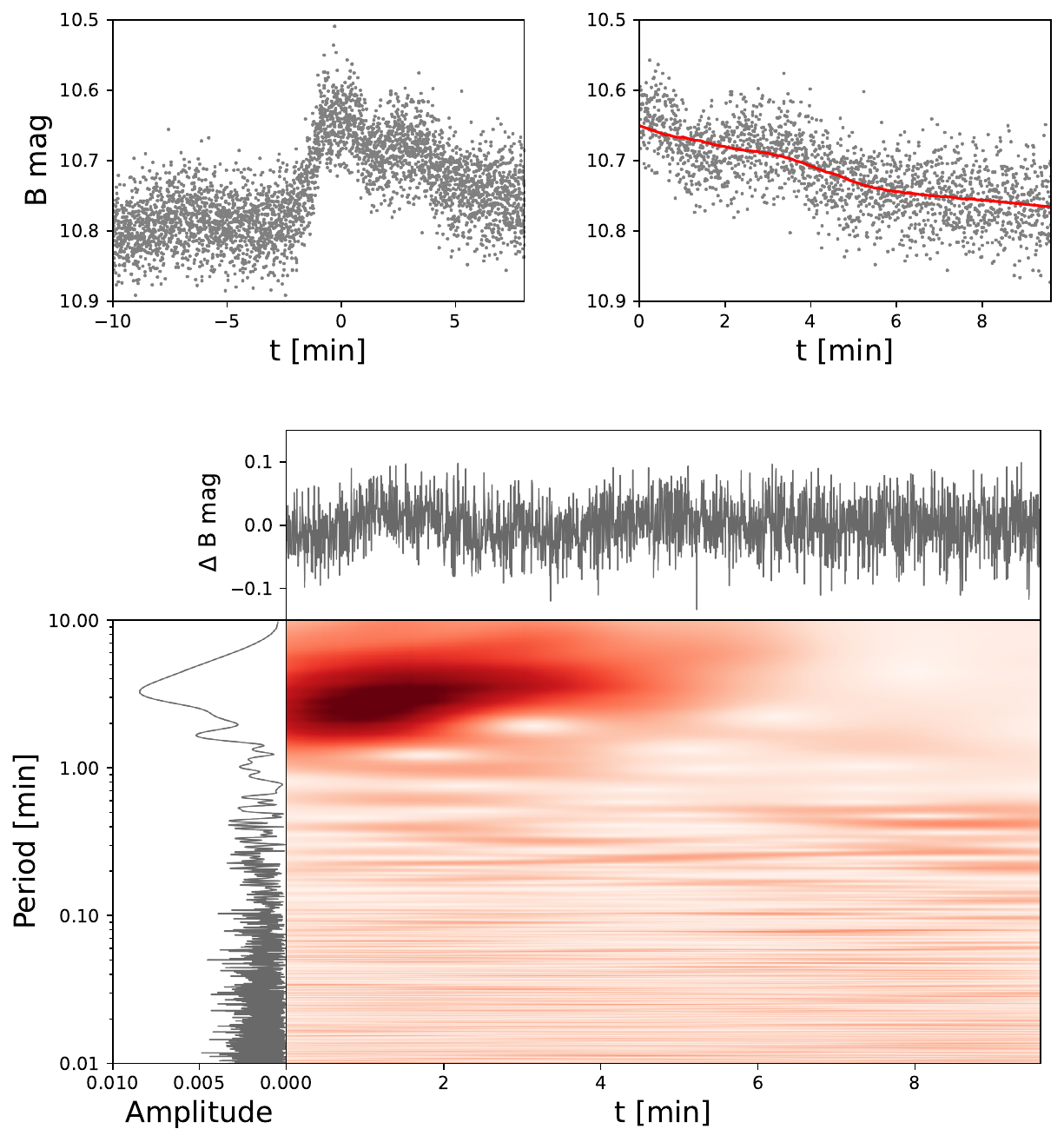}
    \caption{Same as Fig.~\ref{fig:stft_1} for the flare at $\mathrm{HJD}=2458567.44626$. Possible quasi-periodic pulsation is present with a period around 3\,min.}
    \label{fig:stft_2}
\end{figure}

To look for QPPs with periods between 0.6\,s and 10\,min, we checked the windowed Lomb--Scargle transform for the decay phase of each flare. Based on tests with artificial signals injected into real flares, periods within this range can be reliably identified, provided that the signal is sufficiently long (i.e., it lasts for at least a few cycles). We identified two strong signals: one with a period of approximately 1\,min and a significance of $6.1\sigma$, and another with a period of about 3\,min and a significance of $6.8\sigma$. These cases are shown in Figs.~\ref{fig:stft_1} and \ref{fig:stft_2}. No QPP was found with periods on the order of a few seconds.

Most QPP detections reported in the literature have periods similar to those of the candidates identified here. \cite{2024arXiv241207580P} detected 13 QPPs with 1\,s temporal resolution, with periods between 6 and 107\,s. \cite{2022ApJ...926..204H} found 49 QPP candidates with periods ranging from 2 to 36\,min, using 20\,s cadence TESS data. Using more sectors of TESS 20\,s observations, \cite{2025A&A...700A.178J} extended the list with 61 QPPs, with periods between 42 and 193\,s. QPPs on the Sun are observed down to periods of a few seconds \citep{2021SSRv..217...66Z}. \cite{2004AstL...30..319Z} studied a radio flare on AD\,Leo, and found QPP-like signal with a period smoothly changing from 0.5 to 5\,s, and interpreted it in the context of magnetohydrodynamic oscillations.

\subsection{Quantifying flare complexity}

Flare complexity was quantified using the methods outlined in Sect.~\ref{sect:complexity_methods}. To estimate the errors, for each flare we took data segments of the same length as the original, shifted their starting points by a few seconds ($\pm$1\,s, $\pm$3\,s, $\pm$5\,s, $\pm$10\,s, $\pm$30\,s), recalculated the given quantity for each, and finally took the standard deviation of all these values. For the method based on the number of spline terms required for the best GAM fit, the errors were rounded up to the nearest higher integer (since the GAM-based approach can only yield integer values). The complexity values derived from the original, unbinned dataset, together with their estimated errors, are listed in Table~\ref{table:flares}. 

We first use these methods to compare different flares in the sample, and then to examine how temporal binning affects the values for individual flares. All 42 flares were treated as detected at all bin sizes, although in reality smaller flares might not be identifiable at lower temporal resolution.

\begin{figure*}[htb]
    \centering
    \includegraphics[width=2\columnwidth]{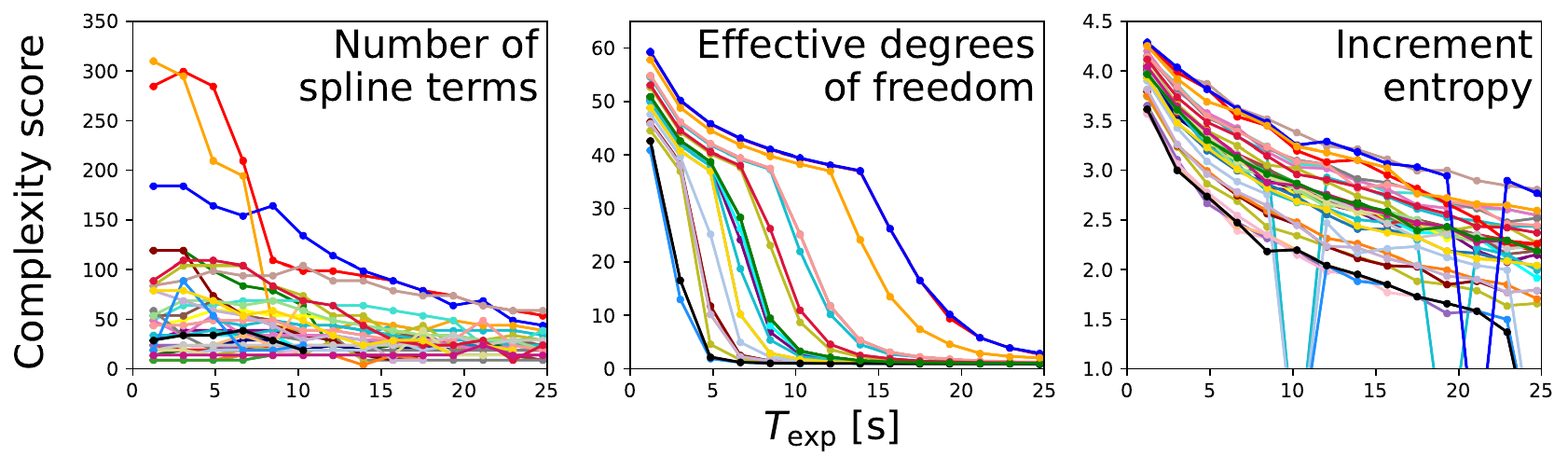}
    \caption{The variation of complexity scores with different bin sizes for the three complexity analysis methods. Each colored line corresponds to a different flare. See the text for details.}
    \label{fig:complexity}
\end{figure*}

As expected, short, low-amplitude flares ranked lowest in complexity, whereas long, high-amplitude flares exhibiting multiple peaks (e.g., the red and orange curves in Fig.~\ref{fig:complexity}, top and bottom panels in Fig.~\ref{fig:multi_peaked_flares}, respectively) or QPPs (e.g., the blue curve in Fig.~\ref{fig:complexity}, top panels in Fig.~\ref{fig:stft_1}) ranked highest. The spline-count method from the GAM fit was particularly effective: unlike the other methods, which were strongly amplitude-dependent and placed the flare at HJD = 2458566.32989 mid-range (see the middle panel of Fig.~\ref{fig:multi_peaked_flares}), it correctly identified it as one of the most complex due to its multiple peaks.

Flare complexity was analyzed as a function of exposure time by assessing ranking performance and changes in complexity values across bin sizes from 1.2\,s to 3\,min. The EDF-based GAM method provided the most consistent flare ranking, maintaining reliable sorting up to 3\,min bins. Increment entropy remained effective up to about 2.5\,min, whereas the spline-count GAM method performed well only at small bin sizes (around 0.5\,min). As expected, the complexity values of individual flares decreased rapidly with binning across all methods (see Fig.~\ref{fig:complexity}). The spline-count GAM method, however, revealed an interesting behavior for double-peaked flares: for the two larger cases (top and bottom panels in Fig.~\ref{fig:multi_peaked_flares}, red and orange curves in Fig.~\ref{fig:complexity}, respectively), the complexity remained nearly constant up to a bin size of approximately 4--5\,s, after which it dropped sharply. Such a plateau is visible for some of the less complex flares as well, suggesting that an integration time of a few seconds does not result in a significant loss of information, and indicating that there is likely no fine structure on shorter timescales in this dataset. However, the high noise level of the light curve also hinders the detections of short-lived and low-amplitude signals.

We can also explore the relationships between the available six parameters for the 42 flares in the sample: amplitude, $t_{1/2}$, bolometric energy, and the three measures of complexity. Based on the Spearman rank correlation coefficient, which measures the strength of general monotonic relationships, there is a strong correlation between the EDF and increment entropy, but the relationship between these parameters and the number of spline terms in the GAM fit is weak. The complexity metrics show only moderate correlations with the amplitude and $t_{1/2}$, indicating that they are indeed independent parameters. The pairwise correlations can be seen in Fig.~\ref{fig:parameter_correlations} in the Appendix.

\section{Conclusions}
Using high-cadence photometry we studied the M dwarf AD\,Leo. From our analysis the following conclusions can be drawn:

\begin{itemize}
\item Using 0.3\,s-cadence $B$-band photometry, in 211\,hours of observations over 49 nights between 2019 and 2025, we identified 42 flare events.
\item The flare energy distribution follows a power-law with a slope of $\alpha = 1.70\pm 0.02$, consistent with earlier results.
\item Three flares showed multi-peaked structures suggesting a complex magnetic energy release.
\item In the case of two events, we found indications of quasi-periodic pulsations in the decay phase of the flares with periods around 1 and 3\,min. No QPP signal was found on the timescale of a few seconds.
\item Based on different complexity measures, we found that the information content of complex flares remains largely preserved up to a binning size of $\approx$4--5\,s.
\end{itemize}

These results can provide practical guidance for planning future observations and space missions capable of fast sampling, e.g., the VISPhot instrument of Ariel \citep{2018ExA....46..135T}. Also, the flare complexity metrics presented here can be used in future studies to describe the temporal morphology of flares in large photometric catalogs, and to rank events based on their complexity.

\section*{Data availability}

The full observed light curve of AD\,Leo is available in electronic form at the CDS via \url{http://cdsweb.u-strasbg.fr/cgi-bin/qcat?J/A+A/...}.

%%%%%%%%%%%%%%%%%%%%%%%%%%%%%%%%%%%
% A C K N O W L E D G E M E N T S
%%%%%%%%%%%%%%%%%%%%%%%%%%%%%%%%%%%

\begin{acknowledgements}
We thank the anonymous reviewer for improving the manuscript with helpful suggestions. This work was supported by the Hungarian National Research, Development and Innovation Office \'Elvonal grant KKP-143986. SB thanks the financial support provided by the undergraduate research assistant program of Konkoly Observatory. This research was also supported by the EKÖP-24 University Excellence Scholarship Program of the Ministry for Culture and Innovation, from the source of the National Research, Development and Innovation Fund. On behalf of the \textit{''Looking for stellar CMEs on different wavelengths''} project we are grateful for the possibility of using HUN-REN Cloud \citep{MTACloud} which helped us achieve the results published in this paper.
\end{acknowledgements}

%%%%%%%%%%%%%%%%%%%%%%%%%%%
% B I B L I O G R A P H Y
%%%%%%%%%%%%%%%%%%%%%%%%%%%

\bibliography{bib}

%%%%%%%%%%%%%%%%%%%
% A P P E N D I X 
%%%%%%%%%%%%%%%%%%%
\begin{appendix}

\onecolumn % to make Appendix C and D fit in a single page

\section{All flares from the sample}

\begin{figure*}[hb!]
    \centering \includegraphics[width=\columnwidth]{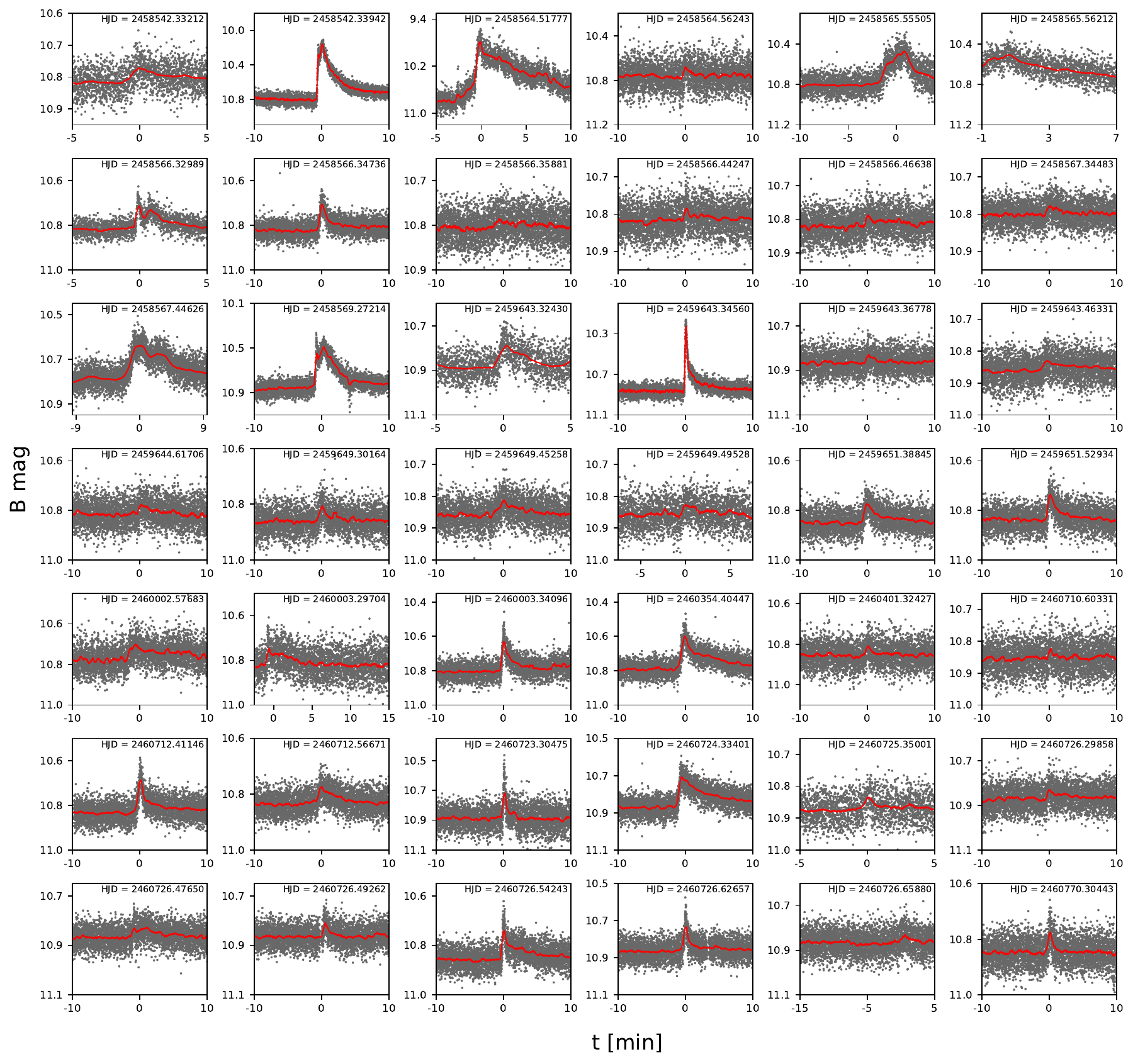}
    \caption{Light curves of all detected flares (grey points), smoothed with a Savitzky--Golay filter (red lines). The HJD of each peak is marked in the corresponding panel.}
    \label{fig:all_flares}
\end{figure*}

\clearpage
\section{Correlations between the measured flare parameters}

\begin{figure*}[hb!]
    \centering \includegraphics[width=\columnwidth]{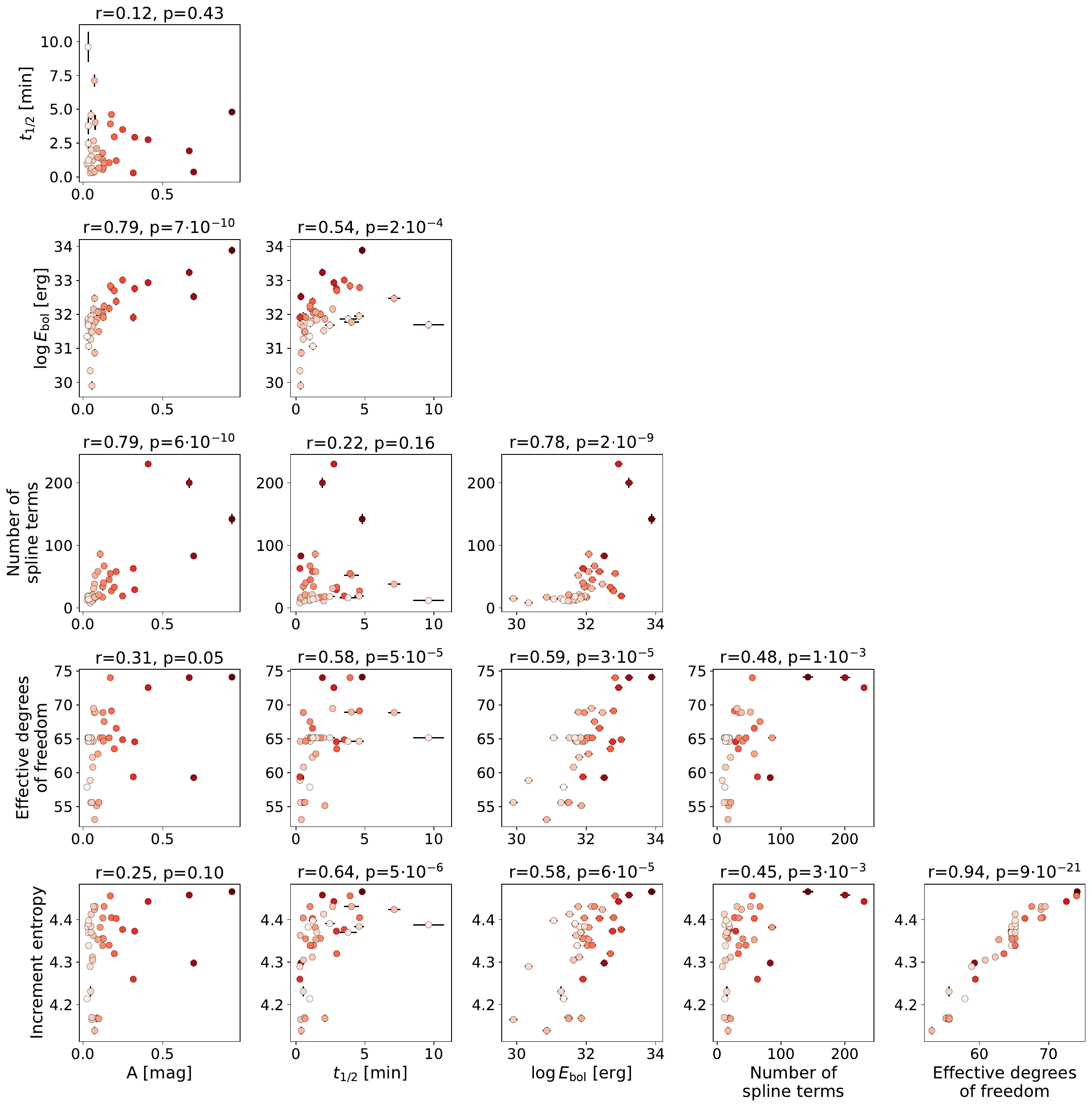}
    \caption{Relationships between the measured flare parameters. The color scale indicates the flare amplitude to make the identification easier across different panels. The Spearman rank correlation coefficients and their $p$-values are shown above each panel.}
    \label{fig:parameter_correlations}
\end{figure*}

\clearpage

\section{Observing log}

\begin{table*}[h!]
\tiny
\caption{Observing log}
\label{table:obs_log}
\centering
\resizebox{0.66\textwidth}{!}{
\begin{tabular}{ccc|ccc|ccc}
\hline
Date & Number of & Noise & Date & Number of & Noise & Date & Number of & Noise \\
 & frames & [mag] & & frames & [mag] & & frames & [mag] \\
\hline
2019-02-23 & 61000 & 0.04 & 2022-03-11 & 99652 & 0.04 & 2024-04-06 & 1931 & 0.08 \\
2019-02-26 & 102000 & 0.05 & 2022-03-12 & 88282 & 0.04 & 2024-04-07 & 30000 & 0.04 \\
2019-02-27 & 101000 & 0.04 & 2022-03-13 & 108216 & 0.08 & 2024-04-09 & 22000 & 0.06 \\
2019-03-21 & 78027 & 0.11 & 2022-03-14 & 19994 & 0.07 & 2024-04-11 & 23600 & 0.06 \\
2019-03-22 & 56000 & 0.06 & 2023-01-12 & 20555 & 0.06 & 2024-04-12 & 25000 & 0.05 \\
2019-03-23 & 80000 & 0.03 & 2023-01-28 & 4779 & 0.05 & 2024-04-13 & 26500 & 0.08 \\
2019-03-24 & 62000 & 0.03 & 2023-01-29 & 51376 & 0.05 & 2024-04-14 & 28000 & 0.08 \\
2019-03-26 & 52899 & 0.05 & 2023-02-24 & 5889 & 0.06 & 2025-02-03 & 91334 & 0.04 \\
2019-03-27 & 50000 & 0.04 & 2023-02-26 & 20108 & 0.05 & 2025-02-04 & 22684 & 0.05 \\
2019-05-18 & 27653 & 0.06 & 2023-02-27 & 50711 & 0.05 & 2025-02-05 & 100000 & 0.04 \\
2019-05-20 & 12149 & 0.31 & 2023-03-01 & 109735 & 0.05 & 2025-02-16 & 83170 & 0.05 \\
2022-03-04 & 53672 & 0.04 & 2024-02-13 & 100000 & 0.04 & 2025-02-17 & 87366 & 0.05 \\
2022-03-05 & 32431 & 0.05 & 2024-02-14 & 11657 & 0.08 & 2025-02-18 & 110000 & 0.05 \\
2022-03-06 & 59619 & 0.05 & 2024-03-30 & 34000 & 0.06 & 2025-02-19 & 120000 & 0.05 \\
2022-03-07 & 17887 & 0.05 & 2024-03-31 & 32619 & 0.06 & 2025-04-04 & 10000 & 0.05 \\
2022-03-08 & 6104 & 0.04 & 2024-04-01 & 28709 & 0.08 \\
2022-03-10 & 93136 & 0.04 & 2024-04-02 & 33936 & 0.06 \\
\hline
\end{tabular}
}
\end{table*}

\section{Flare parameters}

\begin{table*}[h!]
\tiny
\caption{Properties of the detected flares}
\label{table:flares}
\centering

\begin{tabular}{cccccccc}
\hline
Peak time & Amplitude & $t_{1/2}$ & $E_B$ & $E_\mathrm{bol}$ & Number of & Effective degrees & Increment \\
\mbox{[HJD]} & [mag] & [min] & [$10^{32}$\,erg] & [$10^{32}$\,erg] & spline terms & of freedom  & entropy \\
\hline
2458542.33212 & $0.051 \pm 0.003$ & $2.01 \pm 0.19$ & $0.043\pm0.001$ & $0.332\pm0.077$ & $11 \pm 2$ & $65.154 \pm 0.003$ & $4.413 \pm 0.002$ \\
2458542.33942 & $0.666 \pm 0.003$ & $1.92 \pm 0.02$ & $2.225\pm0.008$ & $17.113\pm3.949$ & $200 \pm 8$ & $74.018 \pm 0.001$ & $4.458 \pm 0.001$ \\
2458564.51777 & $0.934 \pm 0.005$ & $4.80 \pm 0.06$ & $9.974\pm0.017$ & $76.727\pm17.706$ & $142 \pm 8$ & $74.093 \pm 0.109$ & $4.466 \pm 0.001$ \\
2458564.56243 & $0.123 \pm 0.009$ & $1.77 \pm 0.25$ & $0.133\pm0.006$ & $1.023\pm0.236$ & $17 \pm 3$ & $65.146 \pm 0.002$ & $4.356 \pm 0.002$ \\
2458565.55505 & $0.324 \pm 0.004$ & $2.93 \pm 0.07$ & $0.748\pm0.005$ & $5.757\pm1.328$ & $29 \pm 3$ & $64.571 \pm 0.270$ & $4.373 \pm 0.004$ \\
2458565.56212 & $0.248 \pm 0.005$ & $3.50 \pm 0.15$ & $1.332\pm0.027$ & $10.245\pm2.364$ & $19 \pm 1$ & $64.879 \pm 0.150$ & $4.377 \pm 0.003$ \\
2458566.32989 & $0.107 \pm 0.003$ & $1.40 \pm 0.06$ & $0.155\pm0.003$ & $1.193\pm0.275$ & $86 \pm 6$ & $65.148 \pm 0.002$ & $4.382 \pm 0.001$ \\
2458566.34736 & $0.123 \pm 0.003$ & $1.27 \pm 0.05$ & $0.142\pm0.002$ & $1.094\pm0.252$ & $34 \pm 3$ & $65.150 \pm 0.002$ & $4.339 \pm 0.001$ \\
2458566.35881 & $0.032 \pm 0.001$ & $9.61 \pm 1.12$ & $0.064\pm0.001$ & $0.494\pm0.114$ & $12 \pm 2$ & $65.154 \pm 0.002$ & $4.388 \pm 0.002$ \\
2458566.44247 & $0.048 \pm 0.005$ & $0.53 \pm 0.08$ & $0.024\pm0.001$ & $0.188\pm0.043$ & $15 \pm 1$ & $55.605 \pm 0.009$ & $4.231 \pm 0.012$ \\
2458566.46638 & $0.030 \pm 0.004$ & $0.89 \pm 0.17$ & $0.087\pm0.001$ & $0.670\pm0.155$ & $19 \pm 2$ & $65.149 \pm 0.002$ & $4.383 \pm 0.005$ \\
2458567.34483 & $0.025 \pm 0.003$ & $1.00 \pm 0.21$ & $0.029\pm0.002$ & $0.224\pm0.052$ & $12 \pm 1$ & $57.896 \pm 0.005$ & $4.214 \pm 0.004$ \\
2458567.44626 & $0.178 \pm 0.002$ & $4.61 \pm 0.10$ & $0.799\pm0.009$ & $6.150\pm1.419$ & $27 \pm 2$ & $69.121 \pm 0.001$ & $4.405 \pm 0.003$ \\
2458569.27214 & $0.408 \pm 0.003$ & $2.75 \pm 0.04$ & $1.116\pm0.004$ & $8.586\pm1.981$ & $230 \pm 4$ & $72.540 \pm 0.001$ & $4.443 \pm 0.001$ \\
2459643.32430 & $0.124 \pm 0.007$ & $0.53 \pm 0.05$ & $0.115\pm0.001$ & $0.888\pm0.205$ & $34 \pm 6$ & $68.877 \pm 0.128$ & $4.404 \pm 0.003$ \\
2459643.34560 & $0.694 \pm 0.005$ & $0.36 \pm 0.01$ & $0.432\pm0.008$ & $3.319\pm0.766$ & $83 \pm 3$ & $59.285 \pm 0.004$ & $4.298 \pm 0.008$ \\
2459643.36778 & $0.064 \pm 0.009$ & $0.32 \pm 0.07$ & $0.067\pm0.001$ & $0.513\pm0.118$ & $13 \pm 2$ & $64.649 \pm 0.002$ & $4.363 \pm 0.003$ \\
2459643.46331 & $0.033 \pm 0.003$ & $3.78 \pm 0.62$ & $0.095\pm0.003$ & $0.729\pm0.168$ & $16 \pm 2$ & $64.640 \pm 0.003$ & $4.370 \pm 0.006$ \\
2459644.61706 & $0.045 \pm 0.015$ & $0.29 \pm 0.15$ & $0.003\pm0.003$ & $0.022\pm0.005$ & $8 \pm 3$ & $58.880 \pm 0.006$ & $4.290 \pm 0.003$ \\
2459649.30164 & $0.059 \pm 0.004$ & $1.19 \pm 0.14$ & $0.082\pm0.002$ & $0.631\pm0.146$ & $21 \pm 2$ & $62.282 \pm 0.004$ & $4.312 \pm 0.002$ \\
2459649.45258 & $0.050 \pm 0.002$ & $4.58 \pm 0.35$ & $0.116\pm0.003$ & $0.889\pm0.205$ & $19 \pm 2$ & $64.654 \pm 0.001$ & $4.384 \pm 0.006$ \\
2459649.49528 & $0.035 \pm 0.005$ & $1.04 \pm 0.22$ & $0.071\pm0.003$ & $0.548\pm0.126$ & $13 \pm 2$ & $64.645 \pm 0.001$ & $4.339 \pm 0.002$ \\
2459651.38845 & $0.086 \pm 0.003$ & $2.10 \pm 0.13$ & $0.095\pm0.004$ & $0.727\pm0.168$ & $18 \pm 2$ & $55.157 \pm 0.007$ & $4.168 \pm 0.009$ \\
2459651.52933 & $0.128 \pm 0.005$ & $0.71 \pm 0.05$ & $0.106\pm0.001$ & $0.813\pm0.188$ & $40 \pm 1$ & $64.661 \pm 0.001$ & $4.355 \pm 0.001$ \\
2460002.57683 & $0.072 \pm 0.003$ & $7.12 \pm 0.45$ & $0.384\pm0.006$ & $2.951\pm0.681$ & $38 \pm 3$ & $68.880 \pm 0.008$ & $4.424 \pm 0.002$ \\
2460003.29704 & $0.076 \pm 0.006$ & $4.02 \pm 0.55$ & $0.077\pm0.001$ & $0.590\pm0.136$ & $52 \pm 2$ & $68.948 \pm 0.003$ & $4.431 \pm 0.002$ \\
2460003.34095 & $0.208 \pm 0.008$ & $1.20 \pm 0.08$ & $0.314\pm0.005$ & $2.412\pm0.557$ & $58 \pm 2$ & $66.562 \pm 0.002$ & $4.403 \pm 0.003$ \\
2460354.40447 & $0.196 \pm 0.003$ & $2.96 \pm 0.07$ & $0.649\pm0.001$ & $4.994\pm1.152$ & $33 \pm 3$ & $63.526 \pm 0.004$ & $4.320 \pm 0.002$ \\
2460401.32427 & $0.061 \pm 0.008$ & $0.55 \pm 0.11$ & $0.056\pm0.001$ & $0.431\pm0.099$ & $13 \pm 1$ & $60.832 \pm 0.004$ & $4.305 \pm 0.002$ \\
2460710.60331 & $0.056 \pm 0.008$ & $0.34 \pm 0.08$ & $0.001\pm0.001$ & $0.008\pm0.002$ & $15 \pm 3$ & $55.632 \pm 0.006$ & $4.165 \pm 0.004$ \\
2460712.41146 & $0.164 \pm 0.004$ & $1.05 \pm 0.04$ & $0.195\pm0.003$ & $1.499\pm0.346$ & $45 \pm 2$ & $65.154 \pm 0.002$ & $4.340 \pm 0.005$ \\
2460712.56671 & $0.066 \pm 0.003$ & $2.66 \pm 0.18$ & $0.186\pm0.003$ & $1.429\pm0.330$ & $31 \pm 3$ & $69.482 \pm 0.002$ & $4.431 \pm 0.002$ \\
2460723.30475 & $0.315 \pm 0.011$ & $0.29 \pm 0.02$ & $0.105\pm0.002$ & $0.809\pm0.187$ & $63 \pm 2$ & $59.406 \pm 0.003$ & $4.260 \pm 0.003$ \\
2460724.33401 & $0.171 \pm 0.002$ & $3.92 \pm 0.09$ & $0.897\pm0.005$ & $6.897\pm1.592$ & $55 \pm 3$ & $74.012 \pm 0.001$ & $4.456 \pm 0.001$ \\
2460725.35001 & $0.055 \pm 0.006$ & $0.62 \pm 0.18$ & $0.040\pm0.002$ & $0.304\pm0.070$ & $11 \pm 1$ & $55.632 \pm 0.004$ & $4.170 \pm 0.006$ \\
2460726.29858 & $0.050 \pm 0.005$ & $1.51 \pm 0.22$ & $0.091\pm0.004$ & $0.697\pm0.161$ & $15 \pm 3$ & $65.154 \pm 0.002$ & $4.370 \pm 0.001$ \\
2460726.47650 & $0.033 \pm 0.003$ & $2.46 \pm 0.35$ & $0.062\pm0.001$ & $0.476\pm0.110$ & $18 \pm 2$ & $65.149 \pm 0.009$ & $4.391 \pm 0.002$ \\
2460726.49285 & $0.073 \pm 0.007$ & $0.39 \pm 0.05$ & $0.009\pm0.001$ & $0.073\pm0.017$ & $17 \pm 1$ & $53.114 \pm 0.002$ & $4.139 \pm 0.009$ \\
2460726.54243 & $0.132 \pm 0.004$ & $1.02 \pm 0.05$ & $0.227\pm0.004$ & $1.743\pm0.402$ & $67 \pm 3$ & $67.534 \pm 0.003$ & $4.431 \pm 0.002$ \\
2460726.62657 & $0.093 \pm 0.005$ & $1.44 \pm 0.12$ & $0.153\pm0.003$ & $1.174\pm0.271$ & $58 \pm 1$ & $62.788 \pm 0.001$ & $4.353 \pm 0.003$ \\
2460726.65880 & $0.035 \pm 0.003$ & $1.23 \pm 0.29$ & $0.015\pm0.001$ & $0.116\pm0.027$ & $14 \pm 3$ & $65.154 \pm 0.006$ & $4.398 \pm 0.001$ \\
2460770.30443 & $0.099 \pm 0.006$ & $0.66 \pm 0.06$ & $0.041\pm0.001$ & $0.314\pm0.072$ & $21 \pm 1$ & $55.645 \pm 0.004$ & $4.167 \pm 0.004$ \\
\hline
\end{tabular}
\tablefoot{The $E_B$ errors only include the uncertainty of ED, while the $E_\mathrm{bol}$ errors also include the uncertainty of the conversion from Eq.~\ref{eq:energy_conversion}.}
\end{table*}

\end{appendix}

\end{document}